\documentclass[10pt,twocolumn,letterpaper]{article}

\usepackage{iccv}              % To produce the CAMERA-READY version
\usepackage[accsupp]{axessibility}  % Improves PDF readability for those with disabilities.

\definecolor{iccvblue}{rgb}{0.21,0.49,0.74}

\usepackage[pagebackref,breaklinks,colorlinks,allcolors=iccvblue]{hyperref}
\usepackage{xcolor}
\definecolor{MyPaleYellow}{HTML}{FFFEBB}
\definecolor{MyPaleGrey}{HTML}{e7e7e7}

\def\paperID{11670} % *** Enter the Paper ID here
\def\confName{ICCV}
\def\confYear{2025}

\title{Optimal Transport for Brain-Image Alignment: Unveiling Redundancy and Synergy in Neural Information Processing}

\author{
    Yang Xiao\textsuperscript{1}\thanks{First Author: yax3417@utulsa.edu} \quad
    Wang Lu\textsuperscript{2}\thanks{Corresponding Author: newlw230630@gmail.com} \quad
    Jie Ji\textsuperscript{3} \quad
    Ruimeng Ye\textsuperscript{1} \quad \\
    Gen Li\textsuperscript{3} \quad
    Xiaolong Ma\textsuperscript{4} \quad
    Bo Hui\textsuperscript{1}\thanks{Corresponding Author: bo-hui@utulsa.edu} \quad
    \\
    \textsuperscript{1}University of Tulsa, \textsuperscript{2}Tsinghua University, \textsuperscript{3}Clemson University,
    \textsuperscript{4}The University of Arizona \\
}

\begin{document}
\maketitle
\begin{abstract}
% Abstract goes here.
The design of artificial neural networks (ANNs) is inspired by the structure of the human brain, and in turn, ANNs offer a potential means to interpret and understand brain signals. Existing methods primarily align brain signals with stimulus signals using Mean Squared Error (MSE), which focuses only on local point-wise alignment and ignores global matching, leading to coarse interpretations and inaccuracies in brain signal decoding. 

In this paper, we address these issues through optimal transport (OT) and theoretically demonstrate why OT provides a more effective alignment strategy than MSE. Specifically, we construct a transport plan between brain voxel embeddings and image embeddings, enabling more precise matching. By controlling the amount of transport, we mitigate the influence of redundant information.
We apply our alignment model directly to the Brain Captioning task by feeding brain signals into a large language model (LLM) instead of images. Our approach achieves state-of-the-art performance across ten evaluation metrics, surpassing the previous best method by an average of 6.11\% in single-subject training and 3.81\% in cross-subject training.
Additionally, we have uncovered several insightful conclusions that align with existing brain research. We unveil the redundancy and synergy of brain information processing through region masking and data dimensionality reduction visualization experiments. We believe our approach paves the way for a more precise understanding of brain signals in the future. The code is available at \url{https://github.com/NKUShaw/OT-Alignment4brain-to-image}.
\end{abstract}
    
\section{Introduction}
\label{sec:intro}
% \sout{Researchers seek to understand the nature of intelligence by studying the structural and functional mechanisms of Brain Neural Network (BNN). Artificial Neural Networks (ANNs), a fundamental component of artificial intelligence, are inspired by the structure and learning principles of BNN~\cite{rosenblatt1961principles}. Neural activities of BNN are related to information processing can be recorded through electrical signals such as electroencephalogram (EEG) and functional Magnetic Resonance Imaging (fMRI)~\cite{ouguz2023introduction}. fMRI captures neural activities indirectly by monitoring fluctuations in blood oxygen levels. The resulting neural activities patterns are then converted into embeddings from pretrained deep learning models, enabling the visualization of internal neural representations and alignment with the other modalities~\cite{scotti2024mindeye2}. Thus we hope to advance research on brain alignment based on deep learning.}
% \lw{
% The structure of the human brain inspired the creation and design of neural networks, and in recent years, research on brain-computer interfaces has become increasingly popular. Researchers have been attempting to use neural networks to understand brain signals and the brain itself. 
% In this paper, we focus on interpreting fMRI signals.
% }

The structure of the human brain has inspired the development of neural networks~\cite{rosenblatt1961principles}, and in recent years, research on brain-computer interfaces has gained increasing attention. Researchers have been exploring ways to leverage neural networks to decode brain signals and better understand brain function. Among various neural recording techniques, functional Magnetic Resonance Imaging (fMRI) provides an indirect measure of neural activity by capturing fluctuations in blood oxygen levels. With advances in deep learning, researchers have attempted to align fMRI signals with neural representations in pre-trained models to enhance brain decoding capabilities~\cite{ouguz2023introduction,scotti2024mindeye2}.
% In this paper, we focus on interpreting fMRI signals by aligning them with stimuli data (e.g., an image signal). 

As the main approach to interpreting brain region signals,
brain-image alignment combines artificial intelligence and neuroscience, allowing us to decipher semantic information from neural activities in response to different types of stimuli~\cite{xia2024umbrae, xia2024dream, zhou2024clip,allen2022massive, wang2024mindbridge, gong2024mindtuner, zou2023re, roth2022natural, luo2024brain, scotti2024mindeye2}.
% \sout{In recent years, brain alignment has emerged as a prominent research direction for combining artificial intelligence and neuroscience, allowing us to decipher semantic information from neural activities in response to kinds of stimuli}~\cite{xia2024umbrae, xia2024dream, zhou2024clip,allen2022massive, wang2024mindbridge, gong2024mindtuner, zou2023re, roth2022natural, luo2024brain, scotti2024mindeye2}. 
% \lw{Brain alignment is currently a mainstream approach for interpreting brain region signals.
% It combines artificial ...
% }
By projecting data in a shared embedding space, information from different modalities can be mapped to the same dimensional space, ensuring semantic consistency~\cite{radford2021learning, li2020unicoder, jia2021scaling, wang2023feature}. By aligning fMRI with the image using deep learning methods, researchers implemented the aligned model for downstream tasks, such as Visual Description~\cite{xia2024umbrae, yang2024neurobind}, Image Generation~\cite{luo2024brain, scotti2024mindeye2, yang2024neurobind}, Object Detection~\cite{xia2024umbrae}, Image Retrieval~\cite{yang2024neurobind}. These advances open up new possibilities for understanding human cognition and developing brain-computer interfaces. 
\begin{figure}
    \centering
    \includegraphics[width=1\linewidth]{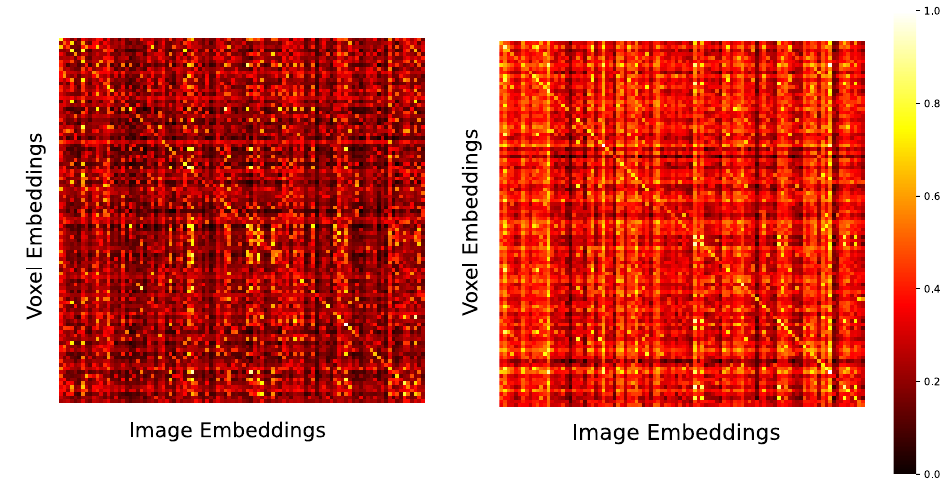}
    \vspace{-4mm}
    \caption{The left is MSE Heatmap and the right is OT Heatmap.}
    \vspace{-4mm}
    \label{fig:heatmap}
\end{figure}
Although brain-image alignment has made significant progress, previous research methods have not considered the differences between brain signal representations and real-world physical signal representations. These methods mainly focus on point-to-point alignment, leading to misalignment and redundancy.
Specifically, most alignment approaches rely on mean squared error (MSE) loss or contrastive InfoNCE loss~\cite{oord2018representation}. 
However, these batch-level loss functions primarily focus on minimizing direct distance metrics, potentially overlooking complex non-linear relationships in cross-modal alignment. As shown in Fig.~\ref{fig:heatmap}, 
MSE can only capture point-wise relationships resulting in highlighted correlations only along the diagonal, while failing to capture global relationships, leading to weak correlations in other areas. Consequently, the underlying mechanisms of alignment remain insufficiently explored~\cite{shiot}. 

To solve the above issues, we analyzed the mathematical principle of MSE alignment. Different modalities can naturally be interpreted as different probability distributions, and aligning these distributions can be formulated as an Optimal Transport (OT) problem ~\cite{stuart2020inverse}. We found that MSE is equivalent to a special case of OT, namely the problem of optimizing a quadratic cost function (Euclidean distance)  when the transport plan is fixed to the identity matrix. However, MSE does not consider whether a better matching strategy exists. Instead, it directly computes the error based on point-wise matching, which can easily lead to local optima. In contrast, general optimal transport builds upon MSE by allowing more flexible matching between different points and considering the alignment of two distributions from a global perspective. Compared to MSE, this approach is more effective in achieving a globally optimal solution. Thus, we propose to use the OT method to achieve brain-image alignment.

We can obtain an optimized transport plan by optimizing the cost matrix, which depends on the distance between one distribution and another. 
Hence, the optimization of the OT problem is equivalent to learning the valid representations in machine learning~\cite{lee2019hierarchical}. 
We constructed a new loss function based on OT (See in Fig.~\ref{fig:overview}). We proved that MSE loss is a special case of OT loss. 
As shown in the OT heatmap (See in Fig.~\ref{fig:heatmap}), compared to MSE, the OT heatmap highlights points outside the diagonal, indicating that it also captures relationships beyond point-wise correspondences. 
Therefore, OT not only considers the point-wise alignment of MSE but also takes into account the impact of the overall distribution. 
By optimizing the loss function, we achieved the \textbf{SOTA} results on downstream tasks much better than MSE loss.

We also verify the existing research conclusions about the brain in cognitive neuroscience from the perspective of deep learning. The brain processes information in both \textbf{redundant} and \textbf{synergistic} ways~\cite{luppi2022synergistic}. Receiving both types of information means that the brain can \textbf{maintain robustness} by utilizing redundant information while achieving \textbf{better processing performance} through synergistic. Redundancy means that increasing the amount of transport information does not always yield optimal outcomes, as excessive information flow may introduce noise or inefficiencies. Therefore, we explored the impact of restricting the mass of transport information on alignment evaluation (See in Fig.~\ref{fig:redundant}). At the same time, the Synergistic nature of brain function is closely tied to its modular organization~\cite{strotzer2009one}. Different brain areas (called Brodmann’s Areas) process different semantic information (See in Fig.~\ref{fig:Schematic})~\cite{brodmann1909vergleichende}. These areas are universally recognized and widely utilized as a standardized anatomical framework for mapping neuropsychological functions within the cerebral cortex~\cite{strotzer2009one}. 
Based on the brain activation, NSD dataset divides the effectively activated cortex into five major brain regions (A brain region consists of multiple brain areas.): 1)  Vision: V1, V2, V3, and V4. 2) Body: EBA, FBA-1, FBA-2, and mTL-bodies ('mid temporal lobe bodies'). 3) Face: OFA, FFA-1, FFA-2, and mTL-faces ('mid temporal lobe faces'). 4) Place: OPA, PPA, and RSC. 5) Word: OWFA, VWFA-1, VWFA-2, mfs-words ('mid fusiform sulcus words'), and mTL-words ('mid temporal lobe words'). On this basis, we not only analyzed the redundancy (See in Tab.~\ref{tab:roi} and Fig.~\ref{fig:dimension_reduction}) and the synergy (See in Tab.~\ref{tab:roi} and Fig.~\ref{fig:dimension_reduction}) between different regions but also examined their differences and importance (See in Fig.~\ref{fig:ROI_importance}).

The following are our contributions:
\begin{itemize}
    \item \textbf{In theory}, We discussed the gap between MSE and OT from the perspective of mathematical derivation and correlation visualization heatmaps and proved that OT Loss can not only learn the point-wise relationship in MSE but also capture the relationship beyond point-wise. 
    \item \textbf{In method}, We established an alignment model between fMRI and images through Optimal Transport methods. The aligned model was then used to test in the pre-trained large language model (LLM), achieving the SOTA in image description by decoding brain fMRI signals. 
    \item \textbf{In analysis}, Our experimental results confirm the presence of both redundancy and synergy in neural information processing. By leveraging PCA and UMAP, we effectively visualized these characteristics, providing deeper insights.
\end{itemize}

% To insert a figure: \input{figs/template}
% Or table: \input{tables/template}
\section{Related Work}
\subsection{Brain-Image Alignment.}
Brain-Image Alignment includes EEG-Image Alignment~\cite{bai2024dreamdiffusion,chen2024mind} and fMRI-Image Alignment~\cite{quan2024psychometry,yang2024brain,xia2024umbrae,qiu2023scratch,scotti2024mindeye2}. 
The goal is to map the neural modality into a common latent space for downstream tasks. Some methods attempt to predict brain responses by taking images as input~\cite{yang2024brain,schrimpf2018brain,zhuang2021unsupervised}. The similarity between ANNs and Brain Neural Network (BNNs) is then quantified based on the correlation between the predicted and actual brain responses, a metric known as the Brain Score~\cite{schrimpf2018brain}. Some methods attempt to use recorded brain responses as the condition to train diffusion models for image generation~\cite{bai2024dreamdiffusion,scotti2024mindeye2,takagi2023high,ozcelik2023natural,scotti2024reconstructing}. Brain Caption aims to generate textual descriptions of a given image based on brain responses rather than the image itself~\cite{takagi2023improving,ferrante2023brain,han2024onellm,xia2024umbrae,shenneuro,ferrante2023brain,mai2023unibrain}. 
\subsection{Optimal Transport.}
Optimal Transport (OT) is a fundamental mathematical framework that seeks to find the most efficient way to transport information or mass from one probability distribution to another~\cite{cuturi2013sinkhorn}. It has been widely studied in probability theory, optimization, and machine learning, with applications spanning image processing, generative modeling, and domain adaptation. 
The Monge Problem~\cite{monge1781memoire} aims to find the best mapping pattern to match two different distributions. Kantorovich formulation~\cite{kantorovich1942transfer} provides the solution to take the Monge problem into a linear programming problem. A key metric derived from Kantorovich’s formulation is the Wasserstein distance, which quantifies the optimal transport cost between two probability distributions. 
The discrete version of the Wasserstein-1 distance was introduced to the image databases as a measure of the minimal effort required to transform one distribution into another by redistributing mass~\cite{rubner1998metric}. 
Many works have successfully applied OT to image classification~\cite{shiot,guo2022learning,shi2024relative}, graph learning~\cite{cheng2025computing,petric2019got,becigneul2020optimal}, transfer learning~\cite{lu2021cross,courty2016optimal,lu2017optimal}.
\subsection{Brodmann's Areas}
Brodmann's Areas~\cite{brodmann1909vergleichende} is universally recognized and widely utilized as a standardized anatomical framework for mapping neuropsychological functions within the cerebral cortex~\cite{strotzer2009one}. These regions, originally defined based on cytoarchitectonic differences, are associated with distinct cognitive, sensory, and motor functions, which work on visual information processing~\cite{ficsek2023cortico}, spatial navigation~\cite{fauzan2015brain}, facial recognition~\cite{rypma2015dopamine}, words decoding~\cite{friederici2017language} or action understanding~\cite{newman2010dissociating}. After visual information enters the human brain, it goes to different partitions and gradually forms representations from primary to advanced levels~\cite{ye2018individual}. The primary visual cortex (V1) is responsible for processing basic visual information~\cite{ye2018individual}, such as orientation and spatial frequency. The Face region includes many areas that are responsible for the initial analysis and integration of facial information, playing a role in emotional processing, and social cognitive activities, and are capable of handling more detailed visual processing~\cite{pitcher2011role,rhodes2009fusiform}. The Word region consists of numerous areas responsible for processing early visual features of text, recognizing words as whole units, and understanding their semantic meanings~\cite{vigneau2005word,dehaene2011unique}. The Place region plays a key role in processing visual geometric features, background information, and spatial memory~\cite{kamps2016occipital,epstein1999parahippocampal,alexander2023rethinking}. The Body Area primarily processes body shape information, motion perception, and the emotional expression of movements~\cite{astafiev2004extrastriate,taylor2007functional}.
\label{sec:related}

\begin{figure*}
    \centering
    \includegraphics[width=1.0\linewidth]{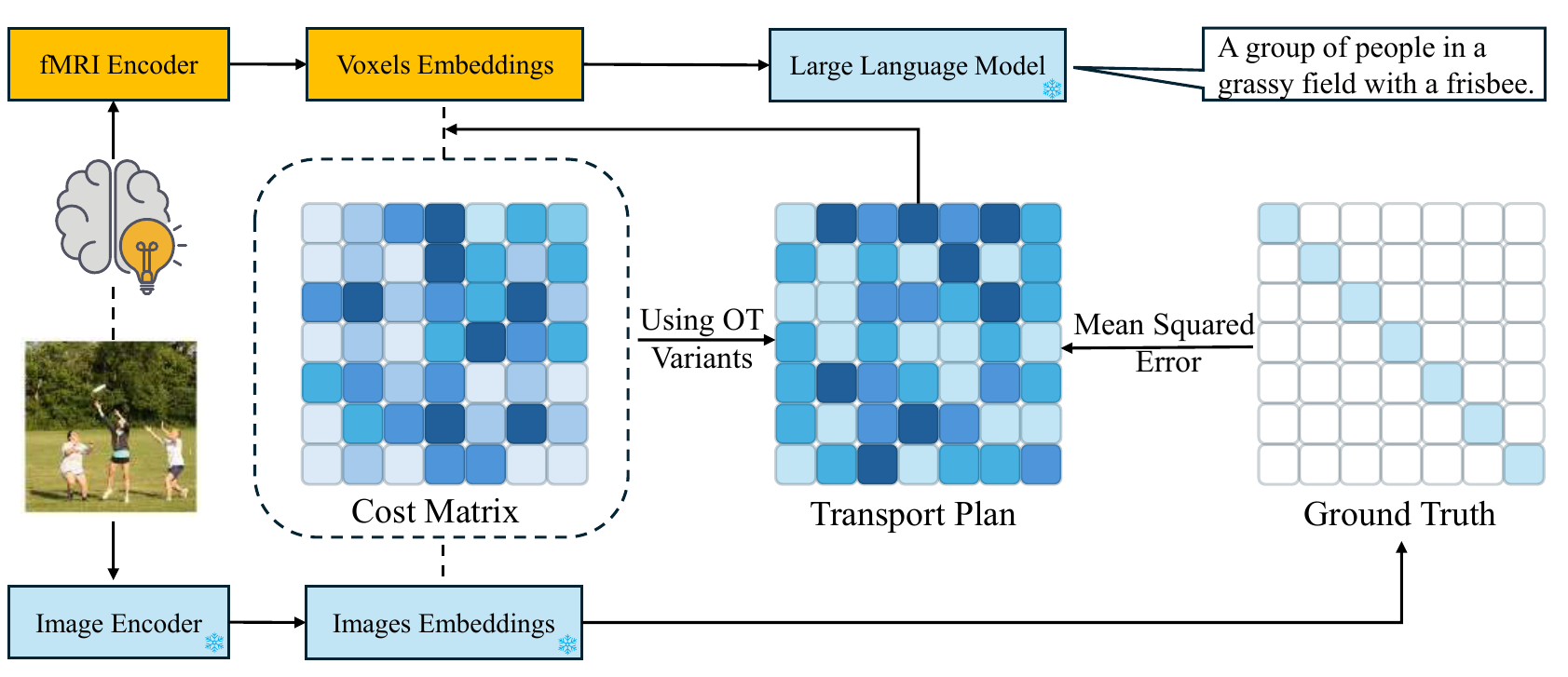}
    \caption{Framework: our OT Loss not only considers the point-wise alignment of MSE but also the global relationships.}
    \label{fig:overview}
\end{figure*}

\begin{figure*}
    \centering
    \includegraphics[width=0.9\linewidth]{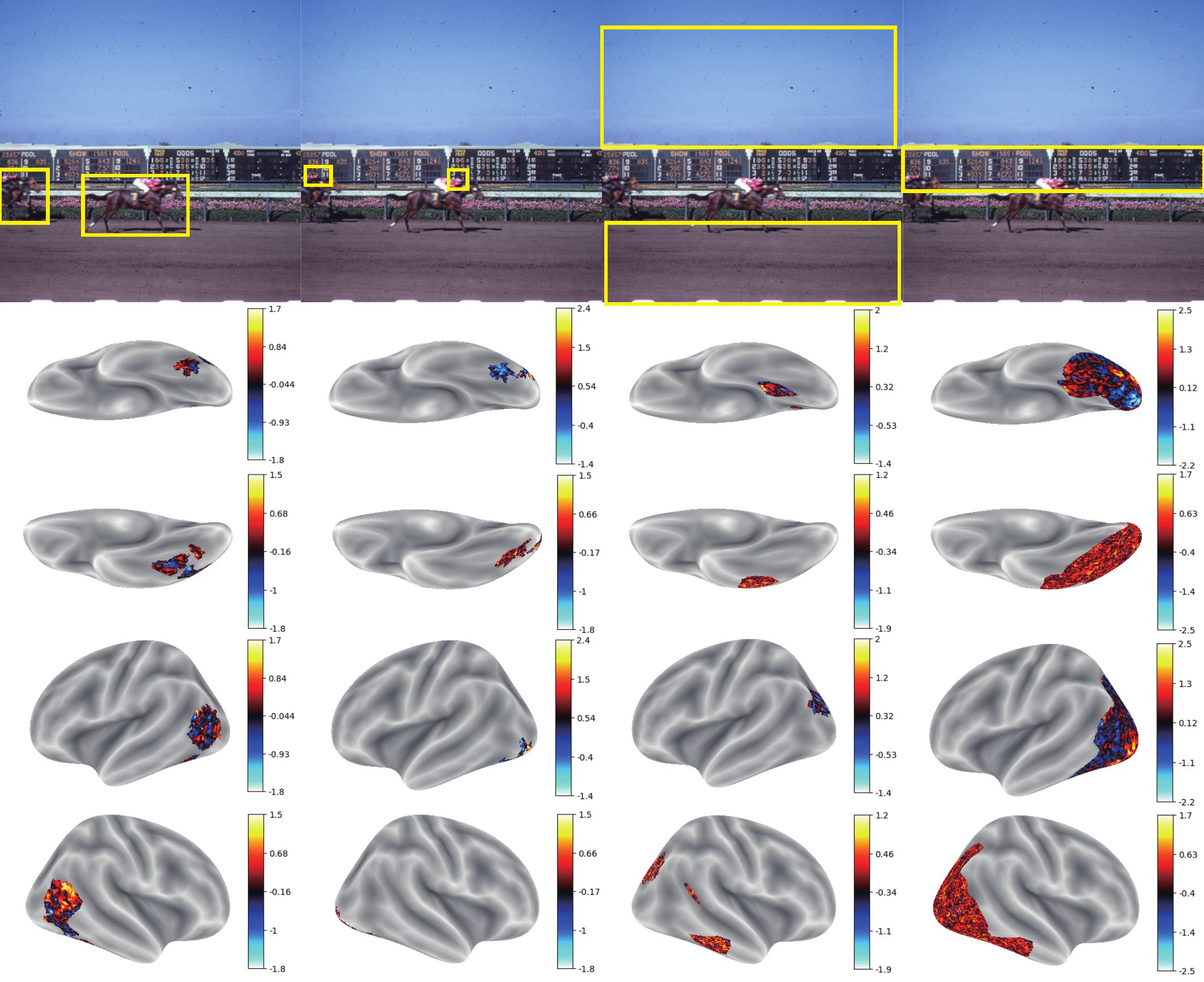}
    \caption{Each row of brain images from upper to down is related to left ventral, right ventral, left lateral, and right lateral.
    Each column from left to right is related to Body, Face, Place, and Word region.}
    \vspace{-4mm}
    \label{fig:Schematic}
\end{figure*}
\section{Problem Setup}
\label{sec:Background}
\subsection{Preliminary}
Let $X$ and $Y$ be metric spaces, $f$ and $g$ be fMRI encoder and image encoder. In this paper, $X$ refers to brain signals and $Y$ refers to images where $x$ is a sample point in $f(X)$ and $y$ is a sample point in $g(Y)$. Let $\mu$ and $\nu$ be probability distributions defined on these spaces:
\begin{equation}
    x\sim \mu, x\in f(X);\text{ }y\sim \nu, y\in g(Y)
\end{equation}

Since $g(Y)$ is a fixed pre-trained weight, our parameter update only involves $f(X)$.

\subsection{Mean Squared Error (MSE)}
A common approach for aligning embeddings is using the Mean Squared Error (MSE) loss. Considering a batch of N fMRI-image pairs $\{(\mathbf{x_i}, \mathbf{y_i})\}_{i}^{N}$
which is defined as:
\begin{equation}
    L_{MSE}(\mathbf{x}, \mathbf{y}) = \frac{1}{N} \sum_{i=1}^{N} \| \mathbf{x_i} - \mathbf{y_i} \|^2
\end{equation}
where $\mathbf{x_i}$ and $\mathbf{y_i}$ represent the corresponding embeddings.
While MSE is a simple and widely used loss function, it has several limitations when applied to aligning distributions:
\begin{itemize}
    \item \textbf{Pointwise comparison}: MSE computes the Euclidean distance between corresponding points but does not consider the overall structure of the distributions.
    \item \textbf{Distribution mismatch}: If $\mu$ and $\nu$ have different support (e.g., discrete vs. continuous distributions), MSE may not provide meaningful alignment.
    \item \textbf{Sensitivity to outliers}: MSE is highly influenced by outliers, which can distort the optimization process.
    \item \textbf{Lack of geometric awareness}: MSE does not account for the underlying geometry of the data space, making it less effective in aligning complex distributions.
\end{itemize}

\section{Method}
\subsection{Framework }
In this section, we will explore how the optimal transport method is applied to our task and how its loss function not only encompasses but also surpasses the MSE loss. While preserving the point-wise training process of MSE, the optimal transport loss additionally captures global relationships, resulting in improved alignment and more powerful model performance.

As shown in Fig.~\ref{fig:overview}, the brain signals are generated by inputting image stimulus. We use fMRI Encoder and a pretrained CLIP Image Encoder to encode both two types of data into the same embedding space. We compute the wasserstein-2 distance of two embeddings as a cost matrix. From the point-wise perspective, we need to compute the MSE distance between pairs of embeddings. From the distribution alignment perspective, we need to measure the global relationships of one distribution on a single point of the other distribution. Ultimately, this forms our transport plan to optimize the loss function and update the model parameters.
\subsection{Optimal Transport}
The Monge Optimal Transport problem~\cite{monge1781memoire} aims to find to a best mapping $T:X\to Y$ to transport the source distribution $\mu$ to target distribution $\nu$ with the least cost $C$:
\begin{equation}
    \min_T\int_{X}C(\mathbf{x}, T(\mathbf{x}))d\mu(\mathbf{x})
\end{equation}
Where \(T\) needs to satisfy the push-forward constraint:
\begin{equation}
    T_{\#} \mu = \nu
\end{equation}
For all measurable sample sets \(S \subseteq Y\), we have
\begin{equation}
    \mu(T^{-1}(S)) = \nu(S)
\end{equation}

\noindent To find a solution of a given OT problem, the Kantorovich formulation~\cite{kantorovich1942transfer} relaxes the mapping constraint in the Monge formulation, allowing the consideration of a joint distribution (transport plan) $\gamma(\mathbf{x}, \mathbf{y})$, thereby transforming the problem into a linear programming problem:
\begin{equation}
    \min_{\gamma\in\Pi(\mu,\nu)}\int_{X\times Y}C(\mathbf{x},\mathbf{y})d\gamma(\mathbf{x},\mathbf{y})
\end{equation}
Where $\Pi(\mu,\nu)$ denotes the set of all joint distributions satisfying the marginal constraints:
\begin{equation}
    \int_Y \gamma(\mathbf{x}, \mathbf{y}) d\mathbf{y} = \mu(\mathbf{x}), \quad \int_X \gamma(\mathbf{x}, \mathbf{y}) d\mathbf{x} = \nu(\mathbf{y})
\end{equation}

The Wasserstein distance addresses both the feasibility issue of Monge and the interpretability issue of Kantorovich. The Wasserstein-p distance is given by:
\begin{equation}
    W_p(\mu,\nu) = (\inf_{\gamma\in\Pi(\mu,\nu)}\int_{X\times Y}\|\mathbf{x}-\mathbf{y}\|^p d\gamma(\mathbf{x},\mathbf{y}))^{\frac{1}{p}}
\end{equation}
Specifically, we use p=2 in our experiments.

Let \( \mathbf{x} \) be the brain voxel embedding and \( \mathbf{y} \) be the image embedding. The Wasserstein-2 distance directly measures the discrepancy between the probability distributions \( \mu \) and \( \nu \), considering both local pointwise differences and global relationships.

\subsection{OT Loss and MSE Loss}
We define the cost matrix:
\begin{equation}
    C_{i,j} = \| \mathbf{x_i} - \mathbf{y_j} \|^2
\end{equation}
where the diagonal of \( C \) represents the Euclidean distance between corresponding points, and the off-diagonal elements account for global relationships.

The size of this cost matrix $C$ is $\mathbb{R}^{N\times M}$, where \( N \) is the number of brain voxel embeddings and \( M \) is the number of image embeddings. (Specially, \( N=M \) in our dataset.) Then, to compute the optimal transport plan, we assume uniform weight distributions:
\begin{equation}
    \mu' = \left( \frac{1}{N}, \frac{1}{N}, \dots, \frac{1}{N} \right), \quad
    \nu' = \left( \frac{1}{N}, \frac{1}{N}, \dots, \frac{1}{N} \right)
\end{equation}
We define $\lambda$ as the mass ratio that controls $m$, the amount of transport:
\begin{equation}
    m = \lambda \min\left( \sum_{i} \mu'_i, \sum_{j} \nu'_j \right).
\end{equation}
The optimal transport plan is obtained by solving:
\begin{equation}
    \gamma^* = \operatorname{argmin}_{\gamma} \sum_{i=1}^{N} \sum_{j=1}^{N} \gamma_{i,j} C_{i,j}
\end{equation}
subject to:
\begin{equation}
    \sum_{i}\gamma_{i,j} \leq \mu'_i, \quad \sum_{j}\gamma_{i,j} \leq \nu'_j, \quad \sum_{i,j}\gamma_{i,j} = m.
\end{equation}
We define the Optimal Transport Loss (OT Loss) as:
\begin{equation}
    L_{OT} = \sum_{i=1}^{N} \sum_{j=1}^{N} \gamma^*_{i,j} C_{i,j}
\end{equation}
which minimizes the Wasserstein-2 transport cost.

By comparing this with the definition of Wasserstein-2 distance:
\begin{equation}
    W_2^2(\mu, \nu) = L_{OT} = \sum_{i=1}^{N} \sum_{j=1}^{N} \gamma^*_{i,j} C_{i,j}.
\end{equation}
To understand the connection between OT Loss and MSE, we rewrite:
% \begin{equation}
% \begin{split}
% L_{OT}&=\frac{1}{B}\sum_{b=1}^{B}( \sum_{i=1}^{N}  \gamma^*_{i,j} C_{i,i}+\sum_{i\neq j}\gamma^*_{i,j} C_{i,j})\\
%       &=\frac{1}{B}\sum_{b=1}^{B}( \sum_{i=1}^{N}  \gamma^*_{i,j} \|\mathbf{x_i}-\mathbf{y_i} \|^2+\sum_{i\neq j}\gamma^*_{i,j} C_{i,j})\\ 
% \end{split}
% \end{equation}
\begin{equation}
\begin{split}
L_{OT}&=\sum_{i=1}^{N}\gamma^*_{i,i}C_{i,i}+\sum_{i=1,}^{N}\sum_{j\neq i}^{N}\gamma^*_{i,j}C_{i,j}\\
      &=\sum_{i=1}^{N}\gamma^*_{i,i}\|\mathbf{x_i}-\mathbf{y_i} \|^2+\sum_{i=1,}^{N}\sum_{j\neq i}^{N}\gamma^*_{i,j}C_{i,j}
\end{split}
\end{equation}
If \( \gamma^* \) is an identity matrix (i.e., mass transport only occurs between corresponding pairs \( (\mathbf{x_i}, \mathbf{y_i}) \)), then:
\begin{equation}
    \begin{split}
        L_{OT}&=\sum_{i=1}^{N} 1\times C_{i,i}+\sum_{i=1}^{N}\sum_{j\neq i}^{N}
        0\times C_{i,j}\\
    \end{split}
\end{equation}
\vspace{-3mm}
\begin{equation}
    L_{OT} \approx L_{MSE}
\end{equation}
which means that MSE is a special case of OT Loss where only diagonal elements contribute. However, in general, OT Loss incorporates a more holistic measure of distribution alignment by considering global transport relationships.

For the model's parameter $\theta$, the parameters updating is:
\vspace{-6mm}
\begin{equation} 
\begin{split}
    \nabla_\theta L_{OT} &= \sum_{i=1}^{N}2 \gamma^{*}_{i,i} \nabla_{\theta}f(X_i)(\mathbf{x_i} - \mathbf{y_i}) \\
    &+ \sum_{i=1}^{N}\sum_{j \neq i}^{N} 2\gamma^{*}_{i,j} \nabla_\theta f(X_i)\frac{\partial C_{i,j}}{\partial \mathbf{x_i}}
\end{split}
\end{equation}
\vspace{-3mm}
\begin{equation} 
\theta \leftarrow \theta - \eta \nabla_\theta L_{OT}
\end{equation}
Where $\eta$ is the learning rate.

\section{Experiment}
\label{sec:experiment}
\subsection{Experiment Setup}
We ran all experiments and visualization in Natural Scenes Dataset (NSD)~\cite{allen2022massive} which contains images from COCO~\cite{lin2014microsoft} and corresponding fMRI signals recorded. Due to spatial distortion and/or head displacement over the course of a scan session, voxels on the edges of the imaged volume may not obtain a full set of data for that session. In pre-processing, such voxels are detected, deemed “invalid”, and are essentially set to 0 for the whole scan session. For the most part, brain voxels of interest are almost always valid. We use the valid voxels as our fMRI part. The dataset spans 8 subjects who were scanned for 30-40 hours (30-40 separate scanning sessions), where each session consisted of viewing 750 images for 3 seconds each. Images were seen 3 times each across the sessions and were unique to each subject, except for a select 1,000 images which were seen by all the subjects. Four subjects (No. 1, 2, 5, and 7) who had completed all the sessions were selected. We choose the test set of subject 1 for evaluation. The model training follows two scenarios: single-subject training (subject 1) and cross-subject training (subjects 1, 2, 5, and 7). In cross-subject training, two subjects are randomly selected in each iteration to compute the OT Loss for parameter updates. 

\subsection{Alignment Experiments}
\noindent\textbf{Image Encoder.} We use the CLIP (Vision Transformer, Large, Patch Size 14, ViT-L/14) model trained by OpenAI to encode image input. This model was trained on a 400M image-text pair dataset constructed by OpenAI, which is larger and more diverse than ImageNet. The extensive training data enables CLIP to generalize across multiple tasks without requiring task-specific fine-tuning. We leverage the pre-trained weights to encode images and align them with corresponding fMRI signals through self-supervised learning, projecting them into an embedding space. Since the pre-trained weights remain frozen, the image embeddings do not change during training, ensuring a stable representation of visual input. This stability facilitates a controlled alignment process between the vision and neural modalities.

\noindent\textbf{fMRI Encoder.} We trained a transformer-based encoder~\cite{jaegle2021perceiver,xia2024umbrae} to map brain voxel signals into the same embedding space as image embeddings for alignment. Transformers are particularly effective in capturing long-range dependencies across voxel signals, making them well-suited for modeling complex neural representations. The fMRI Encoder computes self-attention and cross-attention over voxel-level fMRI signals to capture both intra-modal and cross-modal dependencies. It employs cross-attention mechanisms to condense high-dimensional voxel representations into a latent bottleneck, facilitating dimensionality reduction while preserving essential information. In this process, the key (\(K\)) and value (\(V\)) are derived from projections of the input tokens, while the query (\(Q\)) originates from learnable latent queries. By learning a shared representation, the model aligns fMRI signals with image embeddings, enabling effective cross-modal correspondence.

\begin{table*}[]
    \centering
    \setlength{\tabcolsep}{3pt}
    \resizebox{\textwidth}{!}{
    \begin{tabular}{|c|c|c|c|c|c|c|c|c|c|c|}
    \hline
        Method & BLEU1 & BLEU2 & BLEU3 & BLEU4 & METEOR & ROUGE & CIDEr & SPICE & CLIP-S & RefCLIP-S  \\
        \hline
         \rowcolor{MyPaleGrey}
         Shikra-w/img & 82.38& 69.9& 58.63& 49.66& 35.6& 65.49& 161.43& 27.62& 80.6& 85.92 \\
         SDRecon~\cite{takagi2023high}& 36.21& 17.11& 7.72& 3.43& 10.03& 25.13& 13.83& 5.02& 61.07& 66.36 \\
         OneLLM~\cite{han2024onellm}& 47.04& 26.97& 15.49& 9.51& 13.55& 35.05& 22.99& 6.26& 54.8& 61.28 \\
         UniBrain~\cite{mai2023unibrain}& None & None & None & None & 16.90 & 22.20 & None & None & None & None\\
         BrainCap~\cite{ferrante2023brain}& 55.96& 36.21& 22.7& 14.51& 16.68& 40.69& 41.3& 9.06& 64.31& 69.9 \\
         \cite{shenneuro} & 57.19 & 37.17& 23.78& 15.85& 18.6& 36.67& 49.51& 12.39& 65.49& None \\
         UMBRAE-S1~\cite{xia2024umbrae} & 57.63 & 38.02 & 25.00 &	16.76 &	18.41 &	42.15 &	51.93 &	11.83 &	66.44 &	72.12\\
         UMBRAE~\cite{xia2024umbrae} & 59.44 & 40.48 & 27.66 &	19.03 &	19.45 &	43.71 &	61.06 &	12.79 &	67.78 &	73.54 \\
         UMBRAE(MSE)-S1*& 57.33& 37.69& 24.88& 16.83& 18.23& 41.61& 51.50& 11.86& 66.47& 72.21 \\
         UMBRAE(MSE)*& 59.02& 39.9& 27.1& 18.94& 19.08& 43.41& 58.8& 12.68& 67.87& 73.55 \\
         
         \rowcolor{MyPaleYellow}
         OT-S1 & 63.25& 44.76& 30.96& \textbf{21.72}& \textbf{21.61}& \textbf{47.12}& 71.52& 14.41& \textbf{70.35}& \textbf{75.68}\\
         \rowcolor{MyPaleYellow}
         OT-Cross Subjects & \textbf{64.12} & \textbf{45.18} & \textbf{31.16} & \textbf21.50& 21.42 & 46.83 & \textbf{72.90} & \textbf{14.51} & 69.95 & 75.44 \\
         \hline
         % & & & & & & & & & & \\
    \end{tabular}}
    \caption{Results: all the results cited or we provided are evaluated in the pre-trained shikra.}
    \label{tab:results}
\end{table*}

\begin{table*}
    \centering
    \setlength{\tabcolsep}{1pt}
    \begin{tabular}{|c|c|c|c|c|c|c|c|c|c|c|c|c|c|c|c|c|}
        \hline
        V & P & F & W & B & Other & BLEU1 & BLEU2 & BLEU3 & BLEU4 & METEOR & ROUGE & CIDEr & SPICE & CLIP-S & RefCLIP-S & Valid Voxels \\ \hline
        $\bullet$ & $\bullet$ & $\bullet$ & $\bullet$ & $\bullet$ & $\bullet$ & 63.25& 44.76& 30.96& 21.72& 21.61& 47.12& 71.52& 14.41& 70.35& 75.68 &15724/15724\\
        $\bullet$ & \textopenbullet & \textopenbullet & \textopenbullet & \textopenbullet & \textopenbullet & 43.11& 21.68& 10.23& 5.48& 10.65& 31.64& 10.95& 3.03& 46.24& 52.94& 4657/15724\\
        \textopenbullet & $\bullet$ & \textopenbullet & \textopenbullet & \textopenbullet & \textopenbullet & 48.14& 27.75& 15.09& 8.72& 13.25& 35.23& 23.00& 6.03& 54.25& 60.32& 2252/15724\\
         \textopenbullet & \textopenbullet & $\bullet$ & \textopenbullet & \textopenbullet & \textopenbullet & 44.87& 23.02& 10.89& 5.42& 11.47& 32.58& 12.33& 3.94& 49.10& 55.65& 1042/15724\\
         \textopenbullet & \textopenbullet & \textopenbullet & $\bullet$ & \textopenbullet & \textopenbullet & 42.68& 20.44& 9.16& 4.61& 10.19& 31.45& 8.31& 2.22& 43.99& 50.71& 1736/15724\\
        \textopenbullet & \textopenbullet & \textopenbullet & \textopenbullet & $\bullet$  & \textopenbullet &  53.09& 32.85& 20.38& 12.99& 15.78& 39.05& 37.18& 8.56& 58.23& 64.70& 2934/15724\\
        \textopenbullet & $\bullet$ & $\bullet$ & $\bullet$ & $\bullet$  & \textopenbullet &  59.28& 39.66& 25.91& 16.9& 18.92& 43.52& 55.59& 11.74& 66.13& 71.86& 6706/15724\\
        $\bullet$ & $\bullet$ & $\bullet$ & $\bullet$ & $\bullet$  & \textopenbullet &  61.56& 42.43& 28.65& 19.58& 19.98& 45.05& 62.83& 12.83& 67.34& 73.03& 11051/15724\\
        \hline
    \end{tabular}
    \caption{Region of Interest. V(ison), (P)lace, (F)ace, W(ord), and B(ody).}
    \vspace{-3mm}
    \label{tab:roi}
\end{table*}

\noindent\textbf{Alignment and Brain Captioning.} We aligned the voxel embeddings to image embeddings via OT loss. In evaluation, we use shikra~\cite{chen2023shikra} as the LLM for the brain caption task (Image Description via feeding fMRI instead of image). The LLM will load the voxel embedding instead of image embedding to describe the details of the image.  There are a total of 982 samples per subject.  Ground truth captions are retrieved from COCO~\cite{lin2014microsoft}, and evaluation of inferred captions uses five standard metrics: BLEU-k~\cite{papineni2002bleu},
METEOR~\cite{banerjee2005meteor}, ROUGE-L~\cite{lin2004rouge}, CIDEr~\cite{vedantam2015cider}, and SPICE~\cite{anderson2016spice}, CLIP-S~\cite{radford2021learning}, and RefCLIP-S~\cite{hessel2021clipscore}. During evaluation, we will freeze the weights of the pre-trained LLM. As shown in Tab.~\ref{tab:results}, the first gray row is the result of directly feeding the image into the LLM for evaluation, and this is the gold standard we should try to approach and exceed. Compared with existing methods~\cite{takagi2023high,han2024onellm,mai2023unibrain,ferrante2023brain,shenneuro,xia2024umbrae}, our method achieves SOTA for all metrics in this table. Compared with the best UMBRAE-S1 and the best UMBRAE (Cross Subjects), our method increases average 6.11\% and 3.81\%.

\subsection{Redundant and Synergistic Interactions}
\begin{figure*}[t]
    \centering
    \begin{minipage}{0.33\linewidth}
        \centering
        \includegraphics[width=1.0\linewidth]{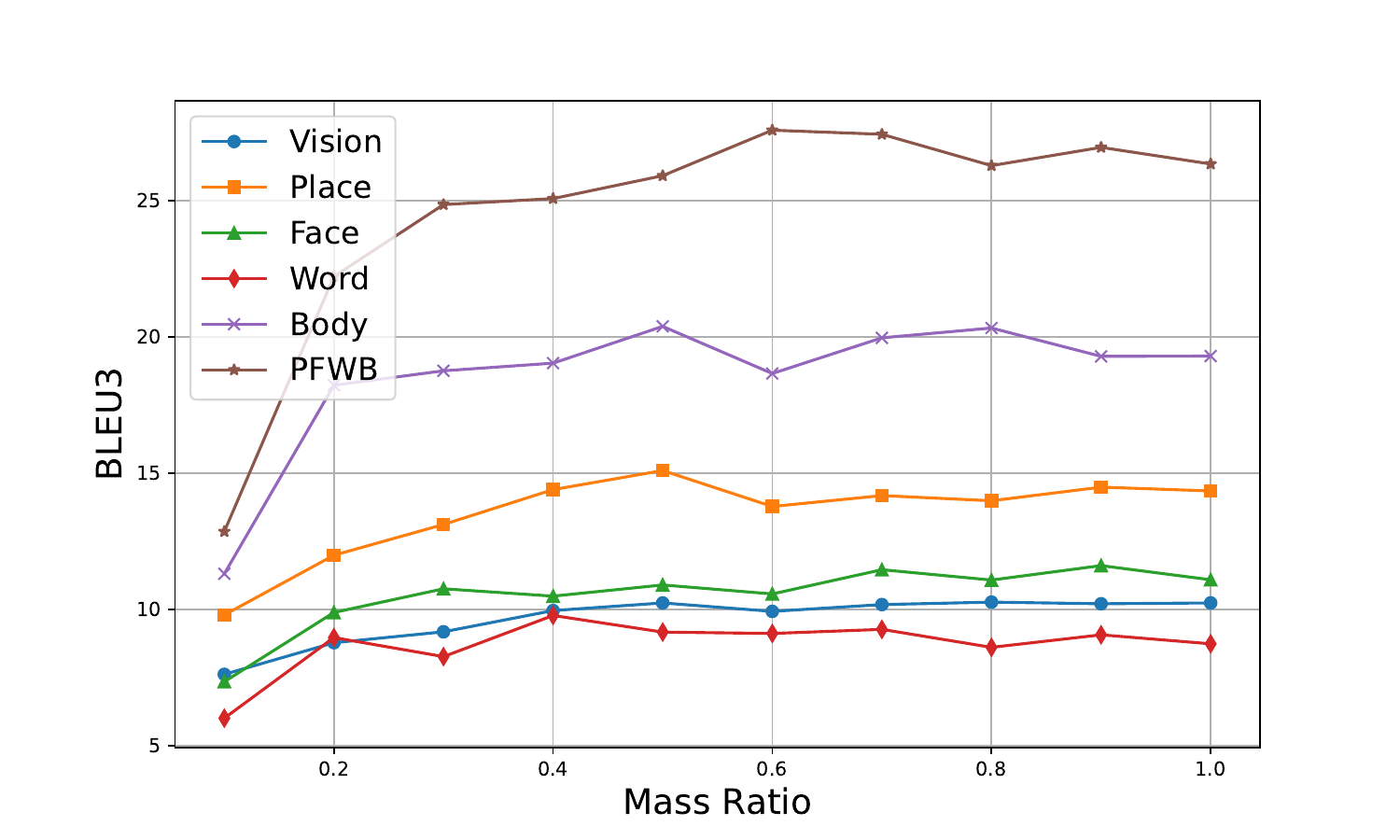}
    \end{minipage}
    \begin{minipage}{0.33\linewidth}
        \centering
        \includegraphics[width=1.0\linewidth]{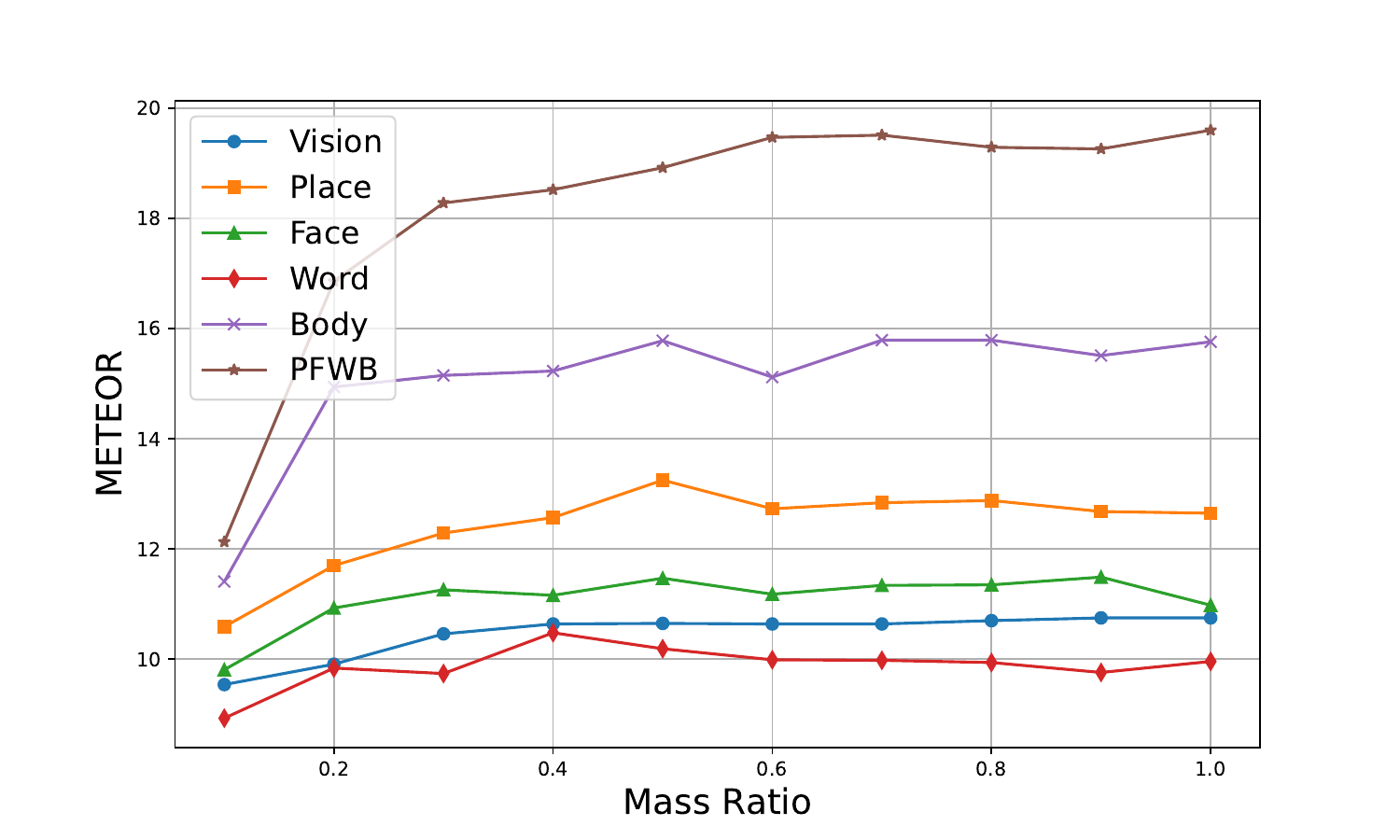}
    \end{minipage}
    \begin{minipage}{0.33\linewidth}
        \centering
        \includegraphics[width=1.0\linewidth]{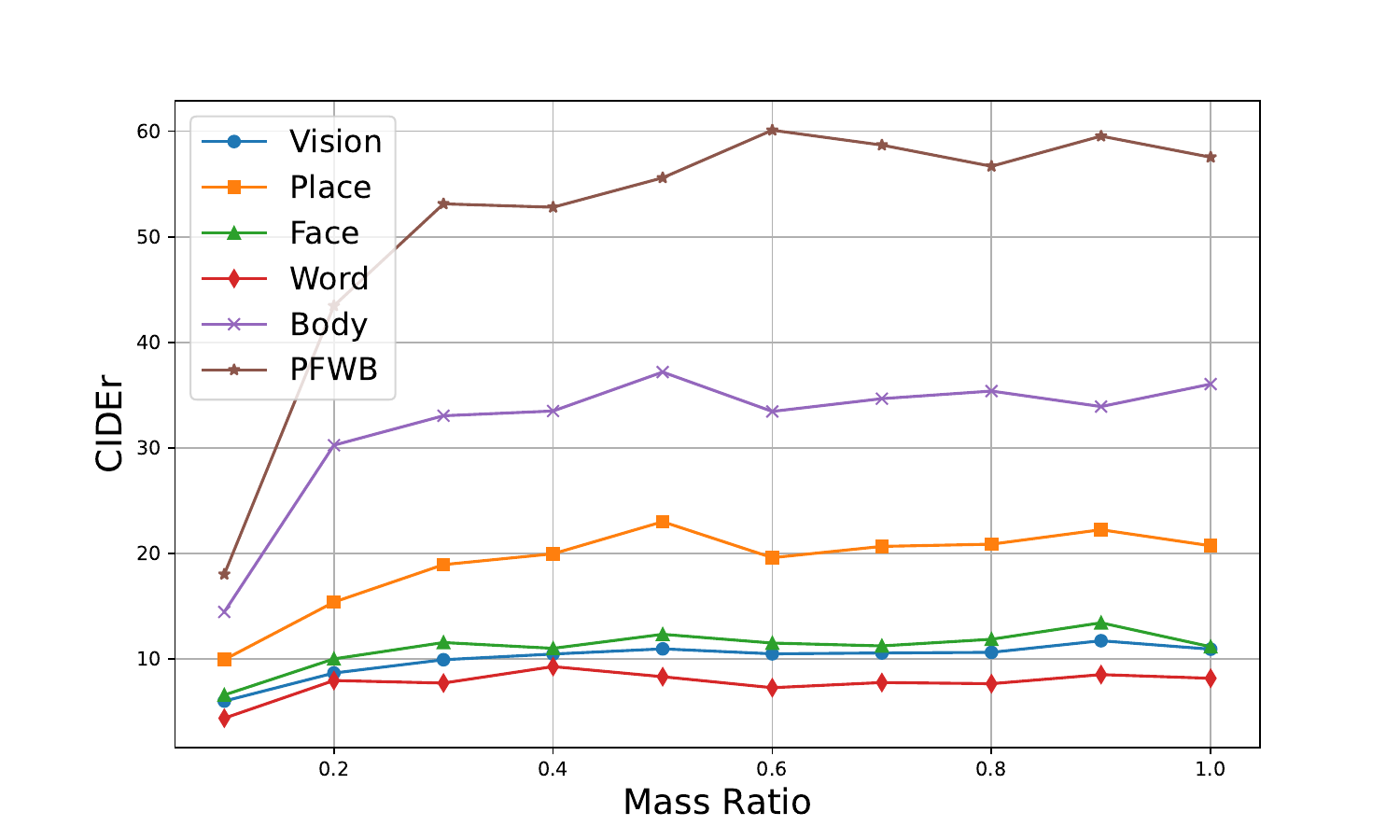}
    \end{minipage}
    \caption{The Redundant and Synergistic of three metrics in different regions}
    \label{fig:redundant}
\end{figure*}

To validate the redundant interactions in brain information processing, we adjusted the mass ratio in the optimal transport process using the hyper-parameter $\lambda$. As shown in Fig.~\ref{fig:redundant}, when the ratio exceeds a certain constant such as 0.3, the model exhibits robust performance. We arbitrarily selected one metric (CIDEr), but in fact, all metrics follow this trend. More detailed results can be seen in table in the appendix A.1. 

\begin{figure}[t]
    \centering
    \includegraphics[width=1.00\linewidth]{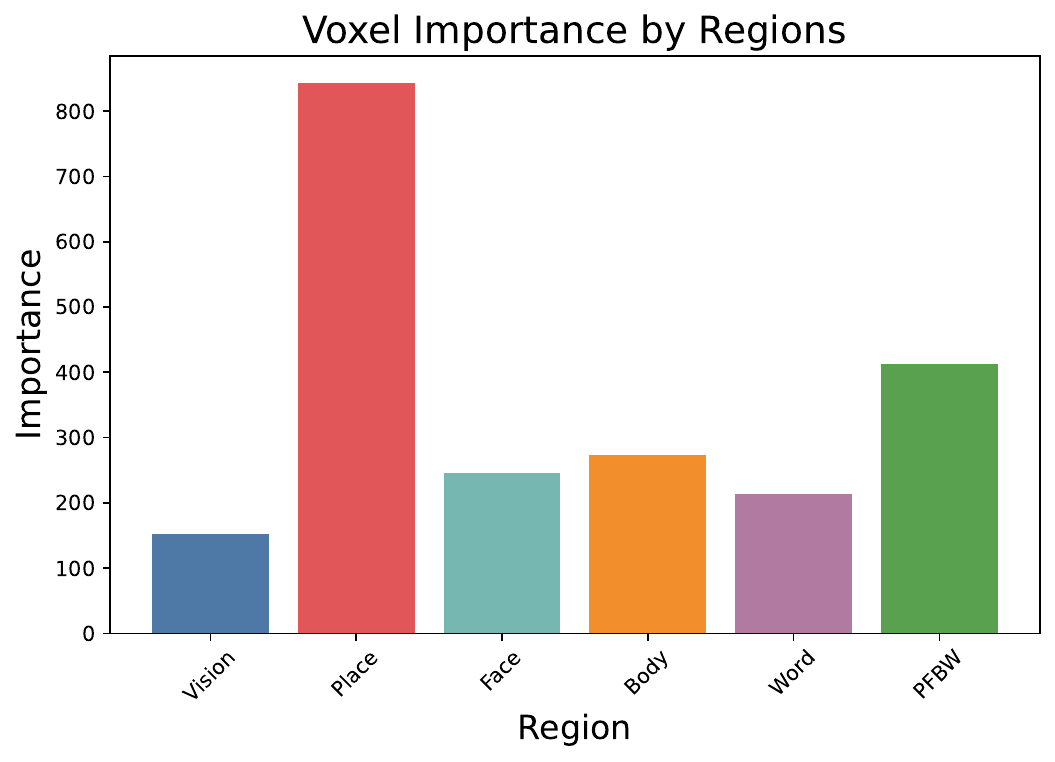}
    \vspace{-2mm}
    \caption{The regions' importance.}
    \vspace{-4mm}
    \label{fig:ROI_importance}
\end{figure}

\begin{figure*}[t]
    \centering
    \begin{minipage}{0.231\linewidth}
        \centering
        \includegraphics[width=\linewidth]{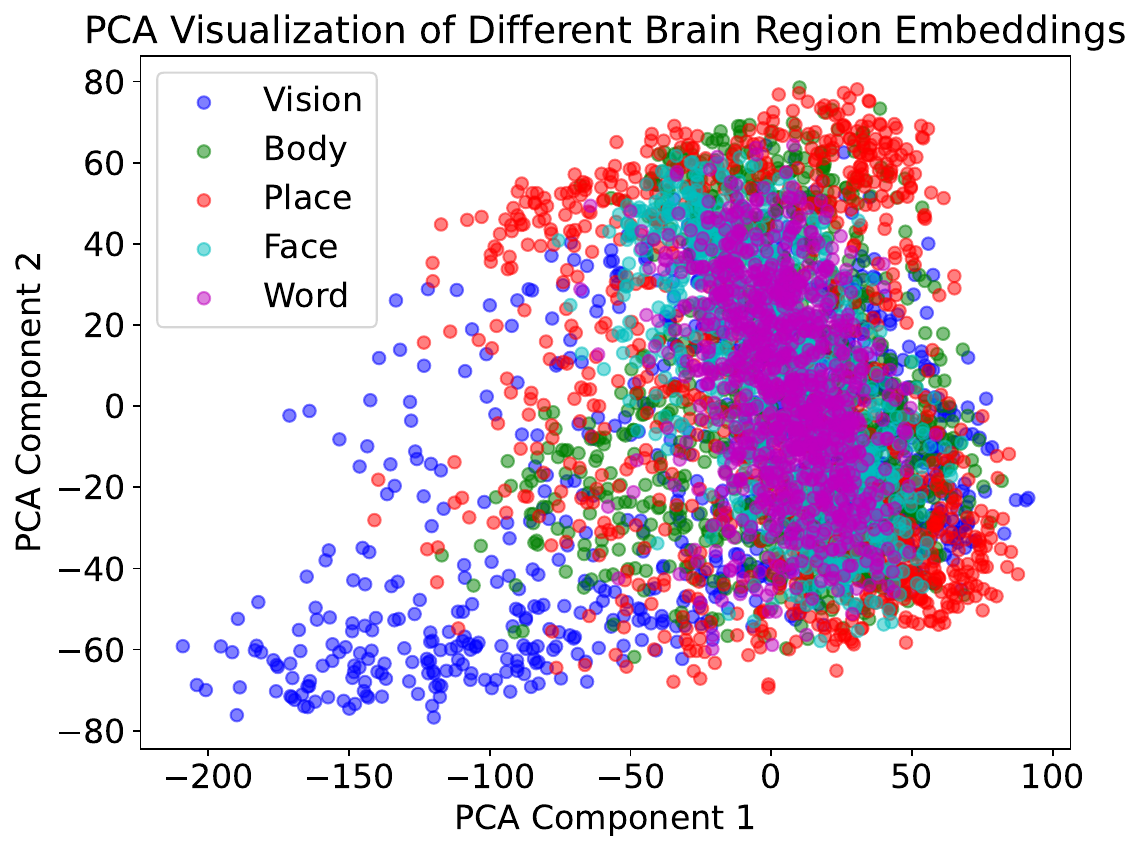}
    \end{minipage}
    \begin{minipage}{0.231\linewidth}
        \centering
        \includegraphics[width=\linewidth]{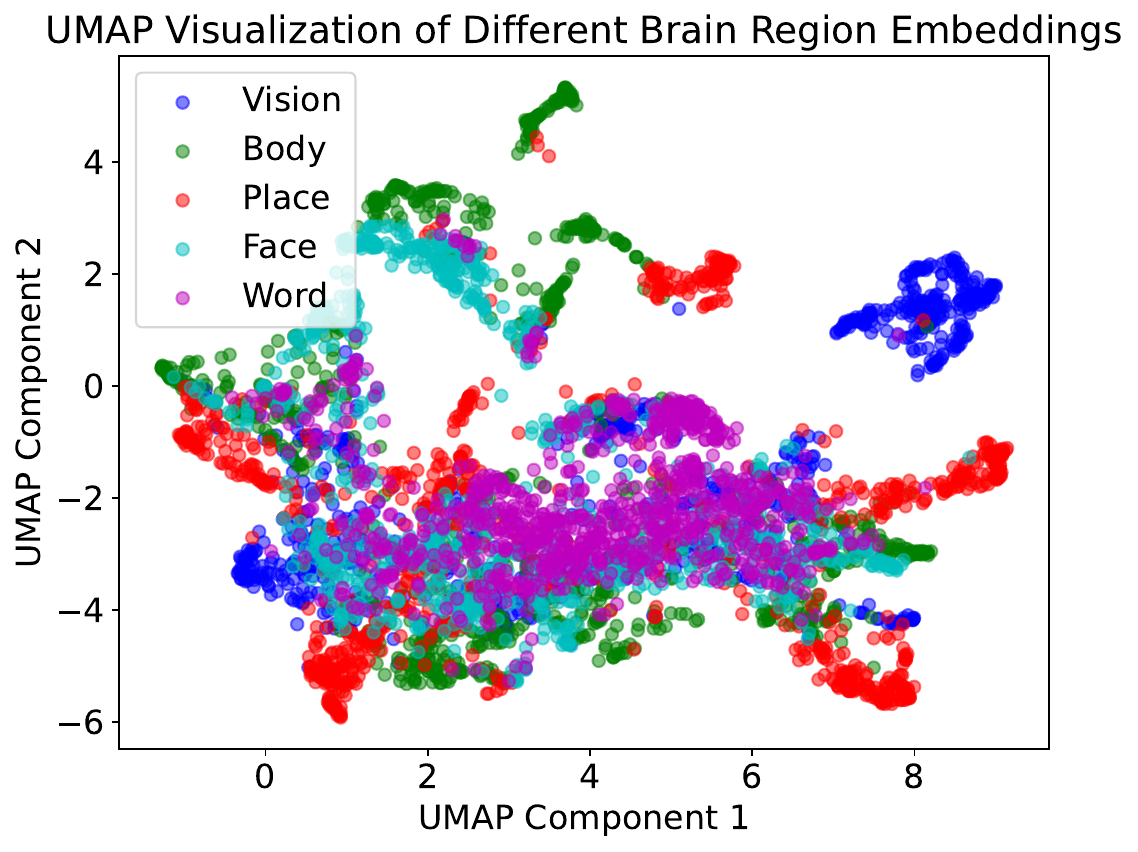}
    \end{minipage}  
    \begin{minipage}{0.231\linewidth}
        \centering
        \includegraphics[width=\linewidth]{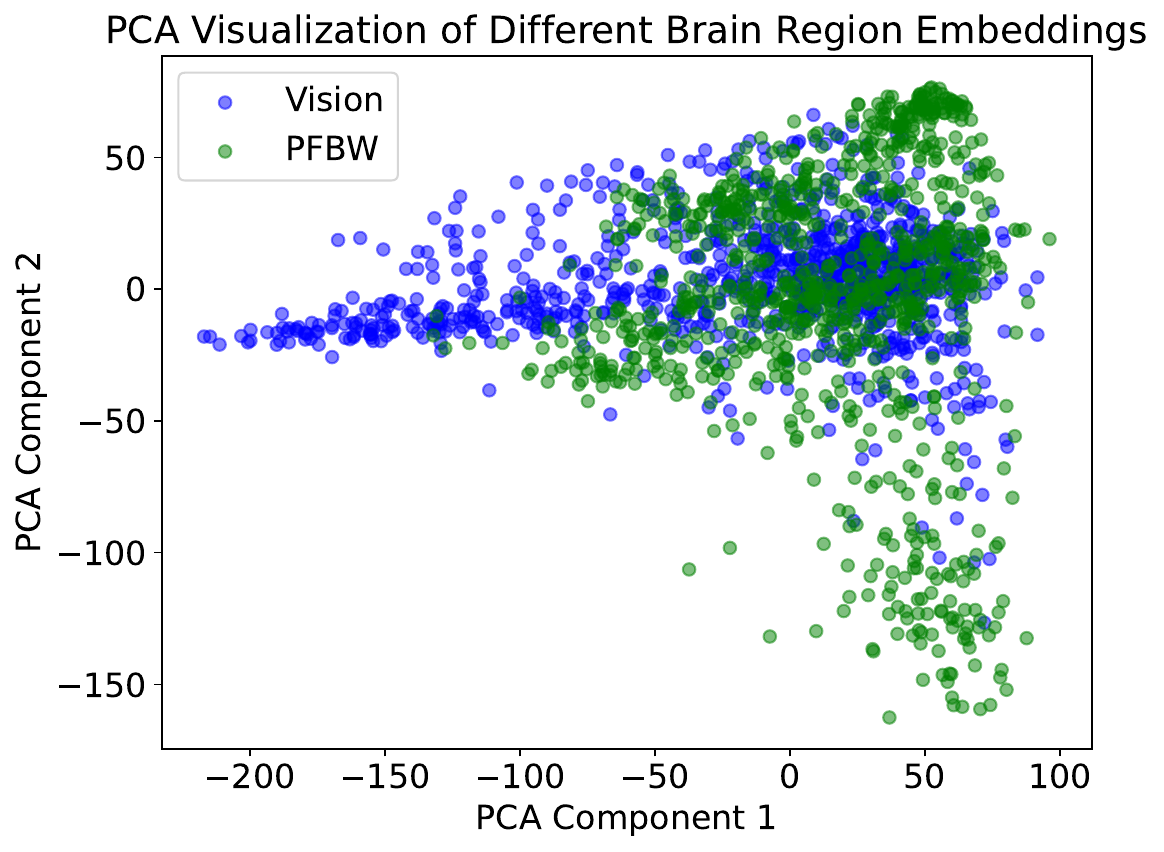}
    \end{minipage}
    \begin{minipage}{0.231\linewidth}
        \centering
        \includegraphics[width=\linewidth]{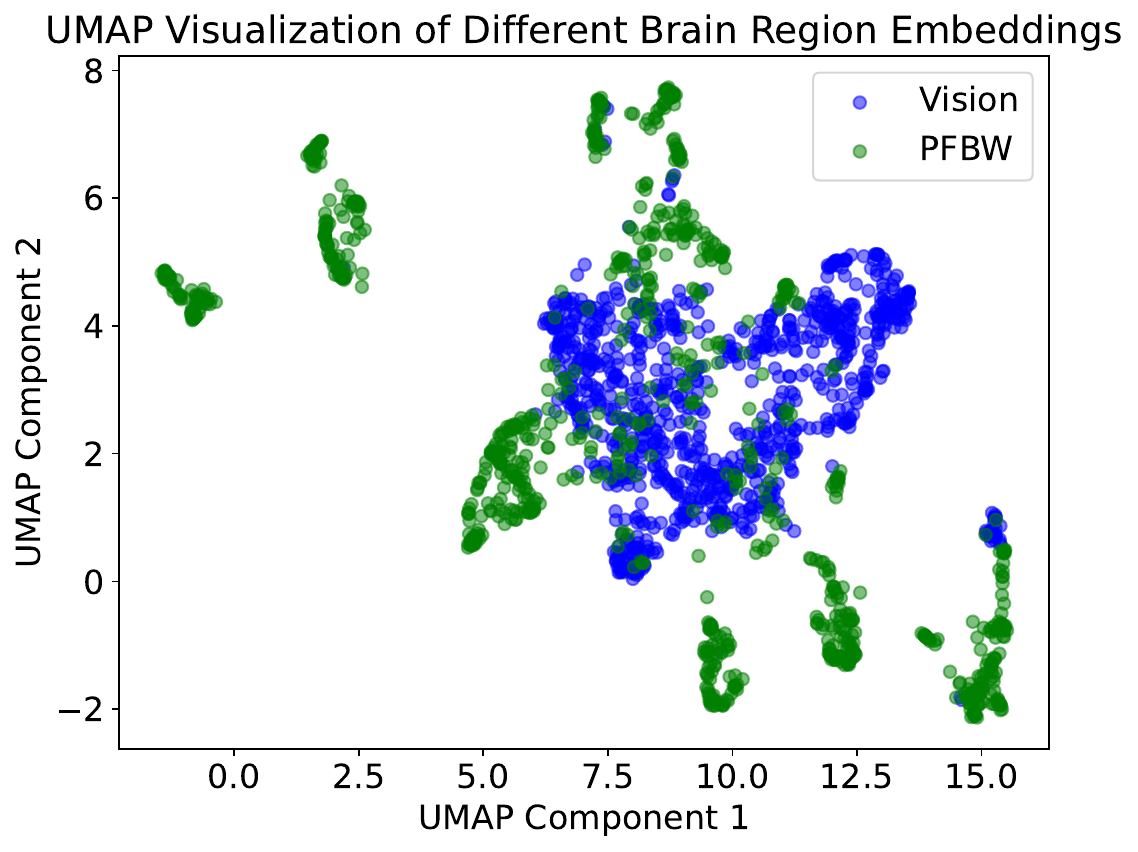}
    \end{minipage}
    \caption{Visualization}
    \label{fig:dimension_reduction}
\end{figure*}
\begin{figure*}[h!]
    \centering
    \includegraphics[width=1.0\linewidth]{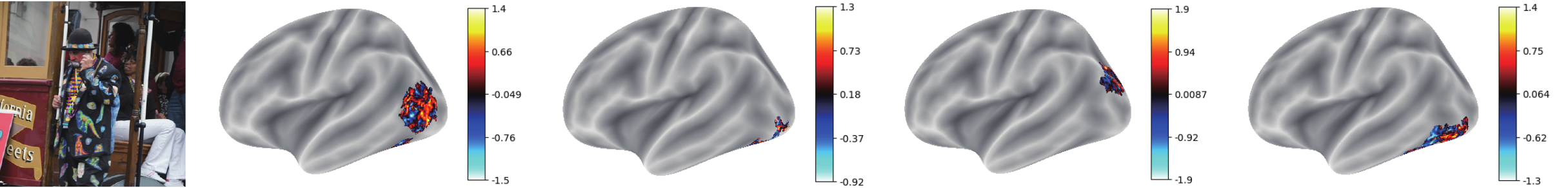}
    \caption{Case study: Ground Truth, Body Region, Face Region, Place Region, Word Region.}
    \label{fig:case 9}
\end{figure*}
We explored synergistic interactions from the perspective of Brodmann’s areas, specifically analyzing the coordination among brain regions related to Vision, Place, Body, Face, and Word. Previous studies have shown that the visual cortex primarily processes low-level information, while certain cortical areas are involved in processing high-level representations. As observed from our experimental results (see Table~\ref{tab:roi}), the performance of a single region approach is suboptimal, and different regions' effects are significant differences while multiple regions working in synergy significantly improve performance.

% \begin{figure}
%     \centering
%     \begin{minipage}{0.75\linewidth}
%         \centering
%         \includegraphics[width=\linewidth]{figs/visualization/brainarea/PCA_Redundancy.pdf}
%     \end{minipage}
%     \begin{minipage}{0.75\linewidth}
%         \centering
%         \includegraphics[width=\linewidth]{figs/visualization/brainarea/UMAP_Redundancy.pdf}
%     \end{minipage}
%     \caption{Dimensionality reduction visualization of Redundant interactions.
%     }
%     \label{fig:dimension_reduction_Redundant}
% \end{figure}

% \begin{figure}
%     \centering
%     \begin{minipage}{0.75\linewidth}
%         \centering
%         \includegraphics[width=\linewidth]{figs/visualization/brainarea/PCA_Synergy.pdf}
%     \end{minipage}
%     \begin{minipage}{0.75\linewidth}
%         \centering
%         \includegraphics[width=\linewidth]{figs/visualization/brainarea/UMAP_Synergy.pdf}
%     \end{minipage}

%     \caption{Dimensionality reduction visualization of Synergistic.}
%     \label{fig:dimension_reduction_synergistic}
% \end{figure}

As shown in Fig.~\ref{fig:Schematic} and~\ref{fig:case 9}, different regions carry different semantic information. 
% Depending on the information contained in the images, the activation situation is also different, 
Based on the information of the images, the activation situation is different, 
for instance, all images contain place information, but not necessarily word information. These semantic attributes influence the activation of different brain regions. To further investigate the contribution of each regions to the alignment task, we employed gradient perturbation analysis and voxel-based weighting for evaluation. The results indicate that the visual region has the lowest importance, whereas the Place region exhibits the highest importance(See in Fig.~\ref{fig:ROI_importance}). This not only corroborates the previous conclusion that the visual cortex primarily processes low-level information but also confirms the ubiquity of Place-related information in images.

We employed PCA~\cite{hotelling1936simplified} and UMAP~\cite{mcinnes2018umap} to visualize embeddings under single and multi-region, aiming to explore their differences. The results indicate that in the visualization of the single region, embeddings appear noticeably interwoven, which may be attributed to redundant interactions. In contrast, under multi-region synergy, the visualization reveals a clear separation between synergistic embeddings and vision embeddings, potentially reflecting the influence of synergistic interactions (See in Fig.~\ref{fig:dimension_reduction}).

In summary, redundancy maintains the robustness of information processing in the brain when external stimuli change, while synergy enables multiple brain regions to collaborate for more efficient information processing. our work made the redundancy and synergy of brain information processing more explicit.

% Redundant interactions demarcate a modular structural-functional backbone in the human brain, ensuring robust sensorimotor input-output channels, whereas synergistic interactions are poised to facilitate high-level cognition through global relationships across different regions, benefiting from diverse patterns of structural connections. 
% In summary, our work made the redundancy and synergy of brain information processing more explicit.

\begin{table}[t]
    \centering
    \resizebox{0.45\textwidth}{!}{
    \begin{tabular}{|c|c|c|c|}
        \hline
         Region& Methods& S-Score(↑)& D-B Index(↓)\\\hline
         5 regions& PCA& -0.0936& 12.7383\\\hline
         V-PFBW& PCA& 0.0261& 5.7202\\\hline
         5 regions& UMAP& -0.0784& 9.1067\\\hline
         V-PFBW& UMAP& 0.1344& 3.6206\\\hline
    \end{tabular}}
    \vspace{-3mm}
    \caption{Quantitative analysis of visualization}
    \label{tab:cluster}
\end{table}
\subsection{Ablation Studies}
Keeping the experimental hyper-parameters and the network architecture unchanged, MSE leads to lower performance compared to OT (See in Tab.~\ref{tab:results}), where UMBRAE(MSE)-S1* is with an average decrease of 6.28\% per evaluation metric, and UMBRAE(MSE)* is with an average decrease of 4.27\% per evaluation metric.
% Keeping the experimental hyper-parameters and the network architecture unchanged, we returned to the MSE loss. The results demonstrate that in single-subject training (See in Tab.~\ref{tab:results}, UMBRAE(MSE)-S1*), all evaluation metrics showed a decrease, with an average decrease of 6.28\% per evaluation metric. In cross-subject training (See in Tab.~\ref{tab:results}, UMBRAE(MSE)*), all metrics also exhibited decrease, with an average decrease of 4.27\% per evaluation metric. 
Then, we calculate the embeddings of the same data set based on two models for correlation analysis. As shown in Fig.~\ref{fig:heatmap}, MSE primarily captures point-wise relationships, resulting in strong correlations only along the diagonal, while failing to account for global relationships. OT extends beyond point-wise correspondences by highlighting correlations beyond the diagonal, demonstrating its ability to capture global structural relationships. Thus, OT not only preserves the point-wise alignment inherent in MSE but also considers the overall distribution, leading to a more comprehensive representation. 
\vspace{-3mm}
\section{Conclusion}
\label{sec:conclusion}
In this paper, we propose an OT-based framework for brain-image alignment, capturing both local and global relationships for better alignment. We first proved why OT exceed MSE. Next, our experiments achieved SOTA in the brain caption task. Finally, we reveal the role of redundancy in neural robustness and synergy of brain regions through Brodmann’s area analysis. This work bridges neuroscience and deep learning, offering new insights.

\section*{Acknowledgement}
This research has been funded in part by the U.S. National Science Foundation grants CRII 2348177.

{
    \small
    \bibliographystyle{ieeenat_fullname}
    \bibliography{main}
}

% \appendix\input{sec/12_appendix}
\end{document}